# Assessing the impact of the coronavirus lockdown on unhappiness, loneliness, and boredom using Google Trends

**Authors:** Abel Brodeur,[1] Andrew E. Clark,[2] Sarah Flèche,[3] Nattavudh Powdthavee[4]

**Affiliations:**

[1] University of Ottawa and IZA. Email: abrodeur@uottawa.ca

[2] Paris School of Economics - CNRS and IZA. Email: Andrew.clark@ens.fr

[3] Aix-Marseille University, CNRS, EHESS, Centrale Marseille, Aix-Marseille School of Economics, Marseille, France. Email: sarah.fleche@univ-amu.fr

[4] Warwick Business School and IZA. Email: nattavudh.powdthavee@wbs.ac.uk

**One Sentence Summary**

The State-imposed lockdowns in Europe and the US have led to a significant increase in Google searches on boredom, loneliness, worry and sadness.

**Abstract**

The COVID-19 pandemic has led many governments to implement lockdowns. While lockdowns may help to contain the spread of the virus, it is possible that substantial damage to population well-being will result. This study relies on Google Trends data and tests whether the lockdowns implemented in Europe and America led to changes in well-being related topic search terms. Using different methods to evaluate the causal effects of lockdown, we find a substantial increase in the search intensity for boredom in Europe and the US. We also found a significant increase in searches for loneliness, worry and sadness, while searches for stress, suicide and divorce on the contrary fell. Our results suggest that people's mental health may have been severely affected by the lockdown.

The COVID-19 pandemic that started in March 2020 has led governments around the world to take unprecedented responses in an attempt to contain the spread of the virus. At the time of writing, some form of State-imposed lockdown has been applied to the residents of many European countries, including France, Italy, Spain and the United Kingdom. Guided by epidemiological models (*1, 2*), the rationale for restricting movement is to save as many lives as possible in the short and medium run. In much of the discourse, the main cost of this confinement has been in terms of the Economy. However, while the cost of lockdown on GDP is considerable, it is possible that there is additional substantial damage to population well-being. Joblessness, social isolation and the lack of freedom, which are some of the by-products of lockdown, are all well-known risk factors for mental health and unhappiness (*3, 4, 5*).

There is on-going research tracking the evolution of well-being during lockdown. For example, a team of researchers at University College London has been collecting mental health and loneliness data of a large sample of adults living in the UK since the day of the lockdown. However, to fully assess how lockdown affects population well-being we also require data from **before the lockdown began**. This is not available in much of the existing research, as most of the lockdown dates were unanticipated. Equally, many standard household surveys that would have been in the field around the lockdown date are likely to have been halted.

We circumvent this problem by analysing data from Google Trends between January 1$^{st}$ 2019 and April 10$^{th}$ 2020 in countries that had introduced a full lockdown by the end of this period. This produces data on nine Western European countries. We also run a comparable analysis at the State level in the US (see Figure S1 and Table S1 for the dates of lockdown). As in previous work using Google Trends to successfully predict disease outbreaks (*6*), tourism flows (*7*) and trading behaviour in financial markets (*8*), we assume that search indicators provide accurate and representative information about the Google Search users' current behaviours and feelings. Furthermore, Google search data shows aggregate measures of search activity in a location (e.g. a State or Country), and is thus less vulnerable to small-sample bias (*9*).

Google Trends supplies an index for search intensity by topic over a requested time period in a geographical area. This is the number of daily searches for the specified topic divided by the maximum number of daily searches for this topic over the time period in that geographical area. This is scaled from zero to 100, where 100 is the day with the most searches for that topic and zero indicates that a given day did not have sufficient search volume for the specific term.

As set out in our pre-analysis plan (OSF; https://osf.io/4ywjc/), we submitted the thirteen following well-being related topic search terms to Google Trends: Boredom, Contentment, Divorce, Impairment, Irritability, Loneliness, Panic, Sadness, Sleep, Stress, Suicide, Well-being and Worry. We have daily data on searches for all of these. This allows us to estimate not only the effect of lockdown on well-being, but also to see whether the intensity of searches changes with the duration of lockdown.

We begin our analysis by comparing the raw data searches pre- and post-lockdown in 2020 to searches pre- and post- the same date in 2019. Here, the lockdown date is the date of lockdown announcement, not the implementation date, as we imagine that the psychological effects of the lockdown would be apparent as soon as the policy is announced to the public (Table S1).

Figure 1 plots daily search activity for three of our search topics: boredom, loneliness and sadness. The results for all topics appear in Figure S2. Searches for boredom in Europe experienced a sharp increase around the announcement date in 2020, while in the US, where the lockdown started later, they began to rise about ten days before the announcement date. This pattern is only seen in 2020, with no sharp changes on the same date in 2019 in either sample. There was a noticeable increase in searches for loneliness in Europe following the lockdown announcement, but not in the US. On the other hand, searches on sadness increased in both samples around one to two weeks after the lockdown.

What explains why certain search topics – such as boredom in the US – registered an uptick in the days before the lockdown announcement? One explanation is that a partial lockdown, which includes school and venue closures, may have already been implemented in these countries (or in some sub-regions within the US State) days before the full lockdown date was announced. It may also reflect people's anticipation of the impending lockdown date based on their observation of areas that had entered lockdown earlier, or the effect of the developing pandemic itself.

To gauge the size of the lockdown effects, Figure 2 reports the Difference-in-Difference (DiD) estimates (Table S2). The top and bottom panels refer respectively to Europe and US. Lockdown is associated with a significant rise in search intensity for boredom in both Europe ($b = 35.8, p < .01$) and the US ($b = 24.0, p < .01$). We also found a significant increase in searches for loneliness ($b_{EU} = 15.9, p < .01; b_{US} = 4.2, p < .1$), worry ($b_{EU} = 12.0, p < .01; b_{US} = 4.1, p < .1$) and sadness ($b_{EU} = 4.6, p < .1; b_{US} = $ 4.1, $p < .01$). The effect size for boredom is large, at two times the standard deviation in Europe and over one standard deviation in the US. The loneliness and worry coefficients are over one half of a standard deviation in Europe, but lower in the US. These can be compared to the estimated standard-deviation effect of 9/11 on mental health of 0.1 to 0.3 (*10*) and depression of 0.5 (*11*) in the US, and an effect on psychological well-being in the UK of 0.07 (*12*). The Boston Bombing had an estimated effect on happiness and net affect of one-third of a standard deviation (*13*).

We also see noticeable drops in stress ($b_{EU} = -12.5, p < .01$; $b_{US} = -5.0, p < .01$), suicide ($b_{EU} = -12.8, p < .01$; $b_{US} = -6.1, p < .01$) and divorce ($b_{EU} = -11.3, p < .01$; $b_{US} = -6.5, p < .01$) in both samples. We found no discernible effect on impairment and a lockdown effect on sleep only in Europe.

Strikingly, the lockdown has a positive effect on the search intensity for the topic of well-being in the US ($b = 13.6, p < .01$) but a negative effect in Europe ($b = -17.3, p < .01$). This could reflect the date at which lockdown was implemented. When we split Europe into early and late lockdowns (this latter group is composed of Ireland, Portugal and the UK), we do indeed find a well-being effect of lockdown in this latter group that is positive. In general, the effect of lockdown on our measures of well-being is often more positive in countries with a later lockdown (Figure S3). Similar conclusions are reached when we use the implementation date as the cut-off (see Table S3). Those entering later lockdowns may be less stressed as they have seen the public-health benefits in countries that entered lockdown earlier.

Is there evidence of adaptation to lockdown? The event study depicted in Figure 3 shows that searches for boredom continued to be higher throughout the lockdown period. Loneliness increased briefly at lockdown announcement before dropping back towards zero in both samples. There was also a gradual increase in sadness after the lockdown. The event studies for all of our variables appear in Figure S4, and the estimated coefficients appear in Table S4.

To test for the immediate structural break caused by the lockdown, we also adopted a regression discontinuity design (RDD), which identifies potential breaks in two parametric series estimated pre- and post-lockdown. As with the DiD estimates, we compare these breaks to those estimated over the same period in 2019 (an RDD-DiD estimation). These estimated breaks are depicted in Figures S5 and S6 for 2020 and 2019, and the estimated coefficients are listed in Table S5. These immediate effects are consistent with those in the event studies: the immediate effect of lockdown was to increase boredom and impairment, reduce panic, but to have little short-run impact on stress, sadness, suicide or worry. DiD and RDD-DiD measure different lockdown effects. The former compares all pre-lockdown observations to all post-lockdown observations, whereas RDD-DiD picks up the immediate effect in the few days around lockdown announcement. This difference is evident in the event studies in Figure 3.

Our results represent the estimated effects of full lockdown announcement. But what about countries such as Germany, the Netherlands and Switzerland where there have only been partial lockdowns (Table S1)? We can include these countries in the lockdown analysis to see if any lockdown is equivalent to full lockdown. Figure S7 compares our main results (in blue) to those for any lockdown (in red). The two figures are similar. We also repeat the same exercise for the US where there was a partial lockdown in some cities and counties before the implementation of a full lockdown at the State level. Figure S8 shows the results when we use the date of the first partial lockdown rather than the date of the full State lockdown. Similar to the European results, there are only small differences in the estimated DiD coefficients. Any announcement of lockdown has substantial effects on a number of measures of well-being.

Our use of Google Trends to assess the well-being impacts of lockdown has important policy implications. Despite the clear message from the government that we should all stay at home to save lives, the evidence of a substantial increase in the search intensity on boredom, sadness, loneliness, and worry post-lockdown suggests that people's mental health has been adversely affected during the first few weeks of lockdown.

We see evidence of mean-reversion in some of these measures, perhaps reflecting individuals' hopes that the lockdown will only be relatively short. Nevertheless, the lockdown effects on boredom and worry have not dissipated over time, and more generally well-being in the first few weeks of lockdown may be only a poor guide to its level after one or two months: we may see accumulated "behavioural fatigue" (*14*) as individuals grow increasingly tired of self-regulating as time passes. To avoid social unrest, it may be necessary to emphasize the health benefits of lockdown (including preparation for testing and tracing after release to avoid a second wave), and make sure that appropriate support is provided to help those struggling the most with lockdown, starting with vulnerable groups (*15*).

**Acknowledgements:**

We would like to thank Mohammad Elfeitori for excellent research assistance.

**Funding:**

Sarah Flèche acknowledges support from the French National Research Agency Grant (ANR-17-EURE-0020).

**Competing interests:**

Authors declare no competing interests.

**Supplementary Materials:**

Material and Methods

Supplementary Tables and Figures

Figures S1-S8

Table S1-S5

**Figures:**

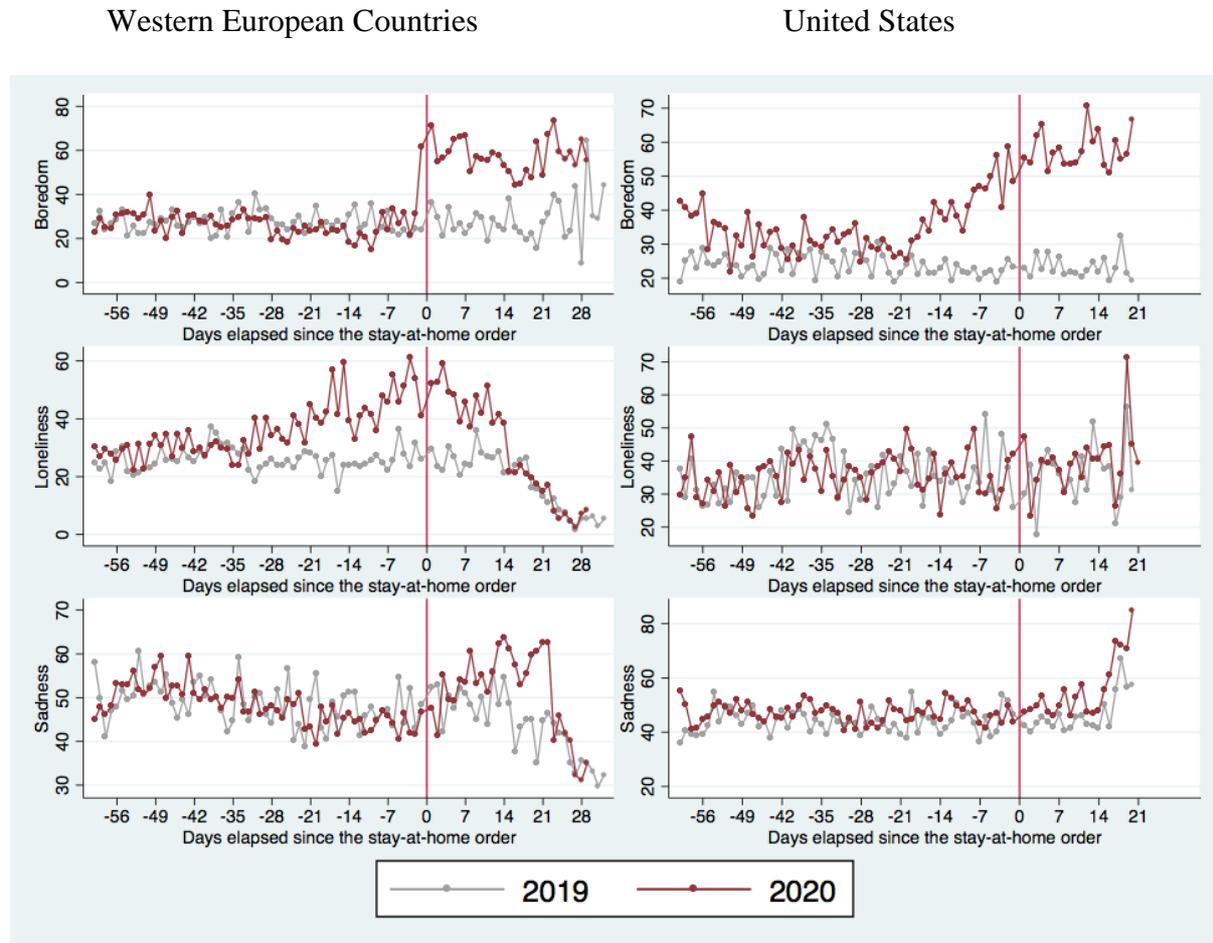

**Fig 1. Google Trends in boredom, loneliness and sadness before and after the stay-at-home orders.** The vertical axis shows the average searches (on a scale from 0 to 100) in the days before (negative values) and after (positive values) the stay-at-home order was announced (set equal to day zero) in 2020 (red dots) and the same date in 2019 (grey dots) for 9 European countries (left) and 42 US States (right). The dots correspond to the raw averages by bins of one day, weighted by the number of inhabitants per country/State. The European countries included are: Austria, Belgium, France, Ireland, Italy, Luxembourg, Portugal, Spain and the UK.

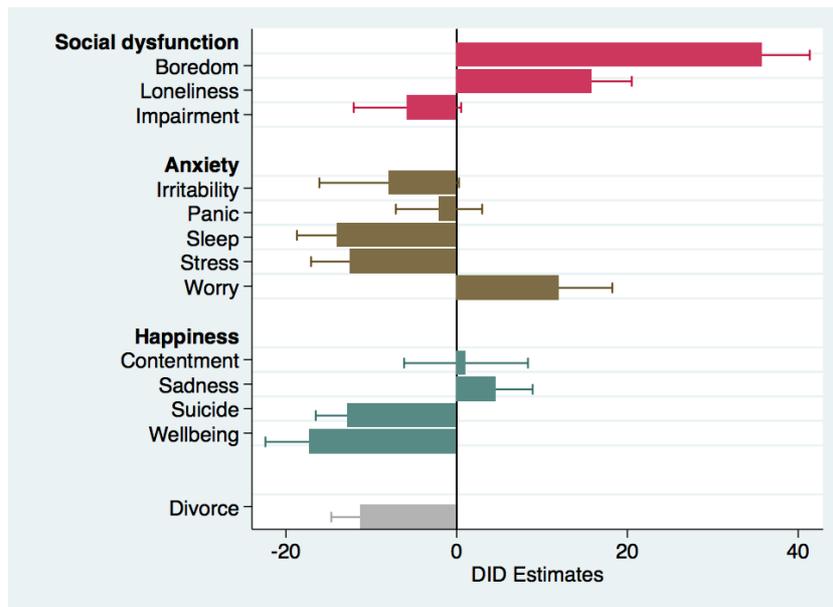

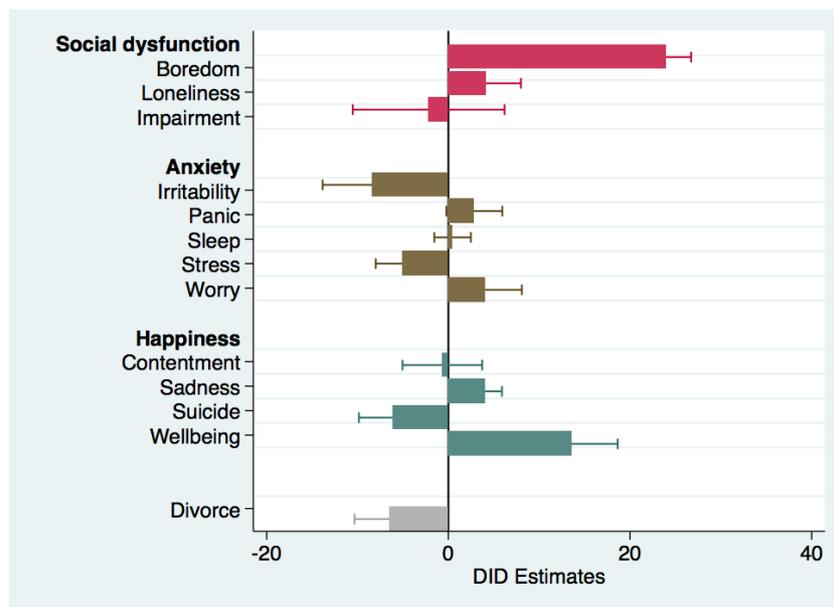

**Fig 2. The effects of the stay-at-home orders on well-being.** Each bar represents Differences-in-Differences estimates using the 2019 period as a counterfactual. All models control for a dummy that takes the value of 1 in the days after the stay-at-home order was announced, as well as country/State, year, week, day of the week fixed effects and the one-day lagged number of new deaths from Covid-19 per million. Weights are applied. Robust standard errors are plotted. Standard errors are clustered at the day level.

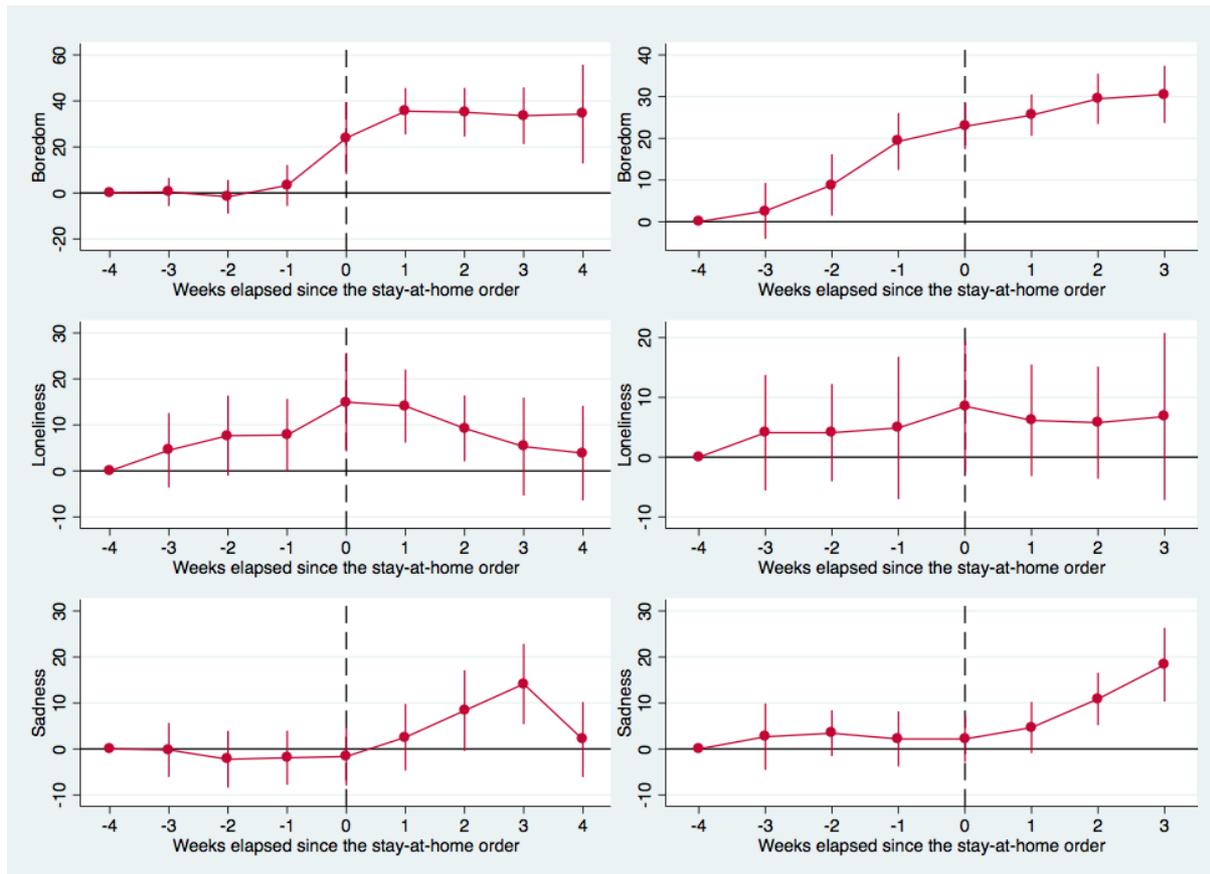

**Fig 3. Duration of the effects of the stay-at-home orders on boredom, loneliness and sadness.** The vertical axis shows event-study estimates using the 2019 period as the counterfactual. The 4[th] week before the stay-at-home-order (in 2019 or 2020) is the reference period. The models include dummies for each week from three weeks before to four weeks after the stay-at-home order. Controls: country/State, year, week, day of the week fixed effects as well as the one-day lagged number of new deaths from Covid-19 per million. Weights are applied. Robust standard errors are plotted. Standard errors are clustered at the day level.